\documentclass[a4paper, 12pt]{article}
\usepackage{graphicx}
\begin{document}

\begin{center}
{\sf {\Large Entanglement generation using silicon wire waveguide }}
\end{center}

\begin{center}
Hiroki Takesue, Yasuhiro Tokura\\
NTT Basic Research Laboratories, NTT Corporation,  3-1 Morinosato Wakamiya, Atsugi, 243-0198, Japan, and\\
CREST, Japan Science and Technology Agency, 4-1-8 Honcho, Kawaguchi, 332-0012, Japan\\
E-mail: htakesue@will.brl.ntt.co.jp
\end{center}

\begin{center}
Hiroshi Fukuda, Tai Tsuchizawa, Toshifumi Watanabe, Koji Yamada, and Sei-ichi Itabashi\\
NTT Microsystem Integration Laboratories, NTT Corporation, 3-1 Morinosato Wakamiya, 
Atsugi, Kanagawa, 243-0198, Japan
\end{center}

\begin{center}
Abstract
\end{center}
We report the first entanglement generation experiment that utilizes a silicon waveguide. Using spontaneous four-wave mixing in a 1.09-cm-long silicon wire waveguide, we generated 1.5-$\mu$m, high-purity time-bin entangled photons without temperature control, and observed a two-photon interference fringe with $>$73\% visibility. 

\newpage

Quantum key distribution (QKD) is now being studied very intensively \cite{gisin}. The key distribution distance of a point-to-point QKD system over fiber recently reached 200 km \cite{np}. However, if we are to construct $>$500 km QKD systems over fiber networks, we need multi-node quantum communication systems such as a quantum repeater \cite{briegel} or a quantum relay \cite{edo,collins}. 

The generation of quantum entanglement in the 1.5-$\mu$m band is a key element for realizing such systems. A candidate for an entanglement source in this band is based on spontaneous parametric downcoversion in periodically poled lithium niobate (PPLN) waveguides \cite{yoshizawa,takesueppln}. In these waveguides, a large group velocity mismatch between a short-wavelength pump and long-wavelength photon pairs induces a walk off between a pump and a photon pair. This walk off causes timing fluctuation of photons, which results in visibility degradation in a quantum interference experiment using two (or more) independent photon-pair sources. Possible solutions for this problem are to use a very short waveguide to suppress the walk off or to use very narrow filters to extend the coherence time of the photon pairs \cite{halder}.  

Spontaneous four-wave mixing (SFWM) in dispersion shifted fiber (DSF) is another promising way to generate 1.5-$\mu$m-band entanglement \cite{li,takesue1,takesue2}. 
In the SFWM process near the zero-dispersion wavelength of the fiber, the four photons that are involved in the nonlinear process have similar frequencies, so refractive index matching and group velocity matching can be achieved simultaneously. As a result, the generated photon pair is close to being Fourier transform limited and thus suitable for quantum interference experiments \cite{takesue3}. This characteristic makes the fiber-based entanglement source promising for long-distance quantum key distribution systems based on quantum relays.    

However, there is a drawback with fiber-based sources, namely noise photons generated by spontaneous Raman scattering (SpRS) \cite{li,takesue1,takesue2}. Although cooling the DSF reduces the number of noise photons \cite{takesue4,lee}, the need for cooling equipment complicates the system and is thus undesirable.

Silicon wire waveguides based on the silicon-on-insulator (SOI) structure are now attracting attention as a nonlinear medium for optical signal processing \cite{tsuchi}. Thanks to the large Kerr nonlinearity of silicon and its very small effective area, the four-wave mixing process occurs very efficiently in such waveguides whose length is of the order of centimeters \cite{fukuda}. In addition, noise photons resulting from SpRS can be significantly suppressed by setting signal and idler frequencies appropriately, since the Raman peak of silicon is 15.6 THz from the pump frequency and has a narrow bandwidth compared with that of DSF \cite{claps}. In 2006, following a theoretical investigation by Lin and Agrawal \cite{lin}, Sharping et al. reported the experimental generation of a correlated photon pair using SFWM in a silicon waveguide \cite{sharping}. However, their experiment has not yet succeeded in generating entangled states. 

In this paper, we report the first experiment to generate entanglement using SFWM in a silicon waveguide. We have generated 1.5-$\mu$m-band time-bin entangled photon pairs \cite{takesue2,brendel} with $>$73\% visibility without subtracting accidental coincidences or detector noises.

\begin{figure}[bht]

\centerline{\includegraphics[width=\linewidth]{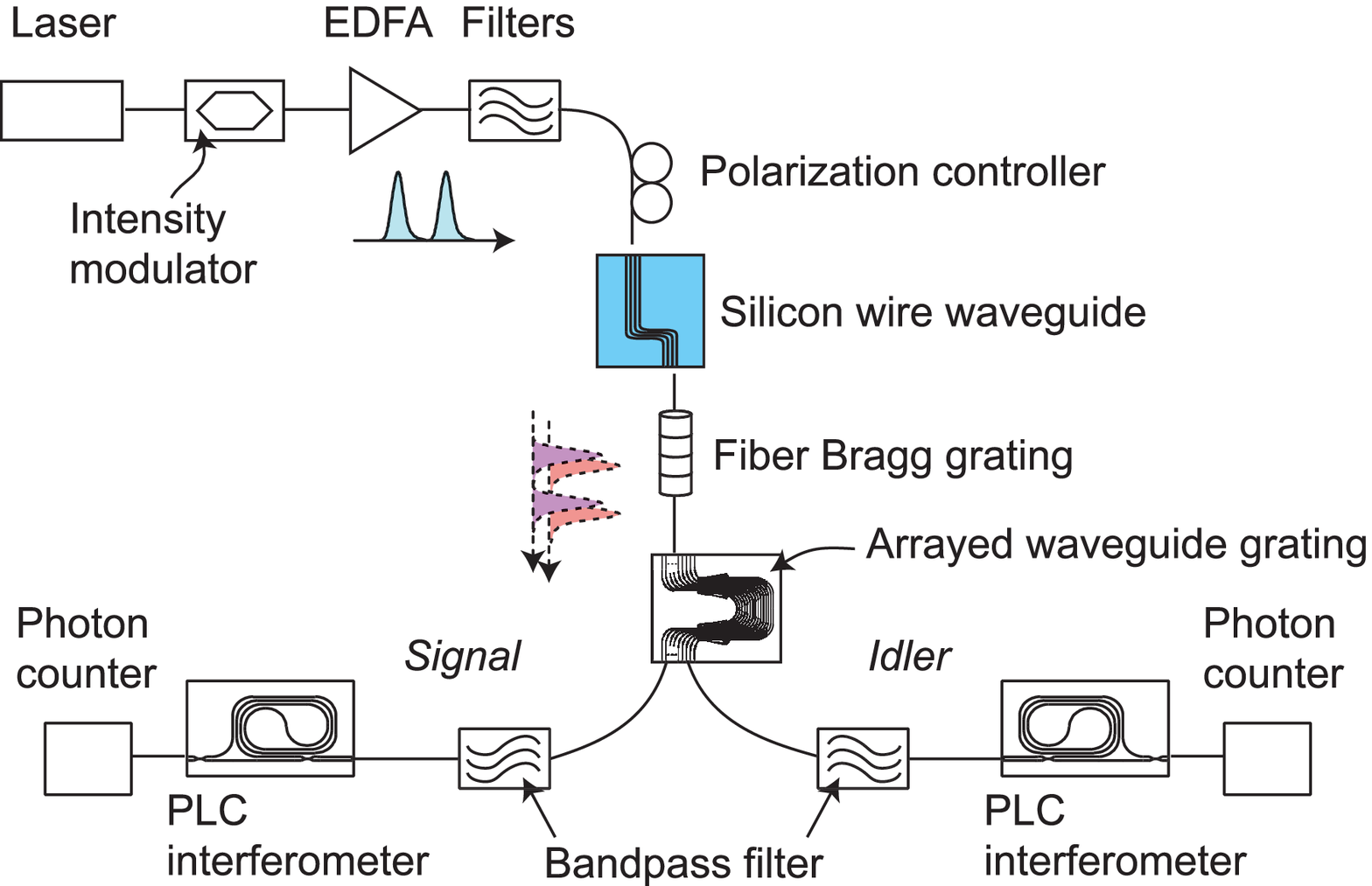}}

\caption{Experimental setup. }
\label{1}
\end{figure}

Figure \ref{1} shows the experimental setup. A 1551.1-nm continuous pump light from an external cavity diode laser was modulated into double pulses with a 100-MHz repetition frequency using an intensity modulator. The pulse width and interval were 90 ps and 1 ns, respectively. The double pulse was amplified by an erbium-doped fiber amplifier (EDFA), transmitted through optical filters to suppress amplified spontaneous emission noise, and injected into a silicon wire waveguide after the polarization state had been adjusted to the TM mode.  The waveguide was fabricated on an SOI wafer with a Si top layer on a 3-$\mu$m SiO$_2$ layer \cite{tsuchi,fukuda}. The Si wire waveguide was 460 nm wide, 220 nm thick, and 1.09 cm long, and required no temperature control. The waveguide loss was 3.1 dB, and the in- and out-coupling losses of the waveguide were estimated to be approximately 4.5 dB. The effective area $A_{eff}$ of the waveguide for the TM mode was calculated to be 0.042 $\mu$m. According to \cite{fukuda}, the nonlinear refractive index $n_s$ was $9 \times 10^{-18}$ m$^2$/W. The nonlinearity coefficient $\gamma$ at a wavelength of 1550 nm is given by
$\gamma = \frac{n_2 \omega}{c A_{eff}}$, 
where $\omega$ and $c$ denote the angular frequency of light and the speed of light in a vacuum, respectively. Using the figures described above, we calculated the $\gamma$ value of the silicon waveguide to be $8.7 \times 10^5$ [1/W/km], which is significantly larger than that of DSF (around 2 [1/W/km]). 

The SFWM process inside the waveguide generated the following time-bin entangled state. 
\begin{equation}
|\Psi\rangle = \frac{1}{\sqrt{2}}\left(|1\rangle_s |1\rangle_i + e^{i \phi}|2\rangle_s |2\rangle_i \right) \label{tb}
\end{equation}
Here, $|k\rangle_x $represents a state in which there is a photon in the $k$th time slot in a mode $x$, signal ($s$) or idler ($i$). $\phi$ is a relative phase term that is equal to $2 \phi_p$, where $\phi_p$ is the phase difference between two pump pulses. The photons output from the waveguide were launched into a fiber Bragg grating to suppress the pump, and input into an arrayed waveguide grating (AWG) to separate the signal and idler photons. The full width at half maximum (FWHM) of the AWG transmittance spectrum was 25 GHz (0.2 nm), which determined the spectral width of the signal and idler photons. The corresponding coherence time of the photon pair was 10.6 ps. The peak frequencies of the signal and idler channels were placed at +400 and -400 GHz from the pump frequency, respectively. The signal and idler photons output from the AWG were passed through optical bandpass filters to further reduce the pump, and launched into 1-bit delayed Mach Zehnder interferometers fabricated using planar lightwave circuit (PLC) technology \cite{takesue2}. The photons from the PLC interferometers were received by photon counters based on InGaAs/InP avalanche photodiodes (Epitaxx EPM239BA), which were operated in a gated mode with a frequency of 5 MHz. The gate signal had a 1.4-ns temporal width and was synchronized with the 100 MHz double pulses. We used the 100 MHz double pulse frequency so that we could easily capture a pulse in a gate with a small change in the gate delay time. The quantum efficiencies and dark count rates per gate of the photon counters were 10\% and $4\times 10^{-5}$ for the signal, and 11\% and $1.2\times 10^{-5}$ for the idler, respectively. The losses for the signal and idler channels, including the effective loss and the out-coupling loss of the silicon waveguide and the insertion losses of the PLC interferometers, were -13.5 and -12.9 dB, respectively. 
The PLC interferometers convert a 
state $|k\rangle_x$ to
$\left(|k\rangle_x + e^{i\theta_x} |k+1\rangle_x \right)/\sqrt{2}$, where 
$\theta_x$ is the phase difference between the two paths of the interferometer for mode $x(=s,i)$, and can be tuned by changing the temperature. 
Then, after passing through the PLC interferometers, the state shown by Eq. 
(\ref{tb}) becomes
$
|\Phi\rangle \to
|1\rangle_s |1\rangle_i +  (e^{i(\theta_s +\theta_i)} +e^{i\phi}) |2\rangle_s 
|2\rangle_i + e^{i(\phi+\theta_s +\theta_i)} |3\rangle_s |3\rangle_i$, where an amplitude term that is common to all product states is omitted for simplicity 
and non-coincident terms are discarded because they do not appear in a coincidence 
measurement. Thus, we can observe two-photon interference at the second time slot while changing the temperature of one of the interferometers.

\begin{figure}[htb]

\centerline{\includegraphics[width=\linewidth]{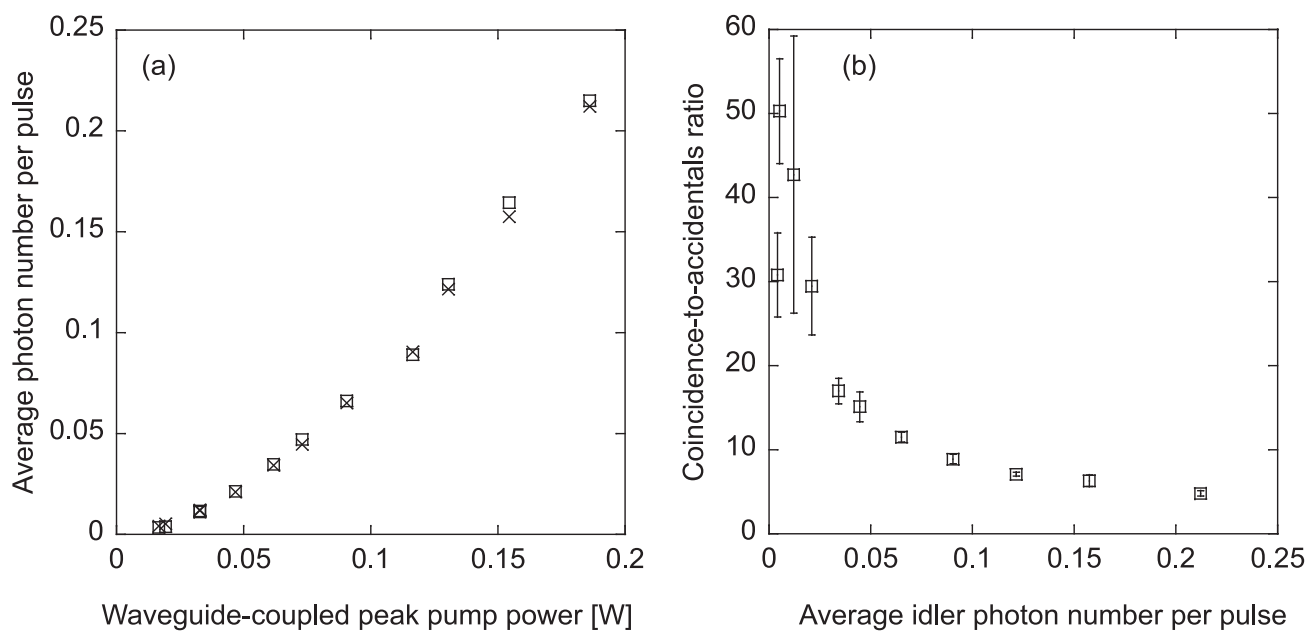}}

\caption{(a) Signal (squares) and idler (x symbols) photon numbers per pulse created in silicon waveguide as function of waveguide-coupled peak pump power, (b) Coincidence-to-accidentals ratio as function of average idler photon number per pulse}
\label{2}
\end{figure}

We first removed the PLC interferometers and measured the characteristics as a photon pair source by using 100-MHz repetition pump pulses with a 90-ps temporal width. Figure \ref{2} (a) shows the obtained average signal (squares) and idler (x symbols) photon numbers per pulse created in the waveguide as a function of waveguide-coupled peak pump power. The waveguide-coupled peak pump power was approximately 120 mW when an average photon-pair number of 0.1 was obtained. 
Based on the theory in \cite{lin}, we estimated $\gamma$ using the experimental data, while neglecting losses caused by two-photon absorption (TPA) and free carrier absorption (FCA). As a result, we obtained $\gamma$ of $3.2 \times 10^5$ [1/W/km], which is approximately one third of the calculated value shown above. The reason for this descrepancy is under investigation. 
Still, the required pump power for the 1.09-cm waveguide was comparable to that needed to generate the same number of photon pairs in our previous photon pair generation experiment using 500 m of DSF with the same pump pulse width and repetition frequency \cite{takesue4}, which proved the high efficiency of the silicon wire waveguide. Figure \ref{2} (b) shows the coincidences-to-accidentals ratio (CAR) as a function of the average idler photon number. The peak CAR was 50.3 at an average idler photon number of 0.005, which is significantly larger than that of around 10 obtained using SFWM in DSF at room temperature \cite{takesue4}. Although this result shows that the number of noise photons observed in the silicon waveguide is at least far smaller than that in a DSF, we need a more thorough investigation to judge whether there are truly no noise photons, which constitutes important future work.

\begin{figure}[t]

\centerline{\includegraphics[width=.8\linewidth]{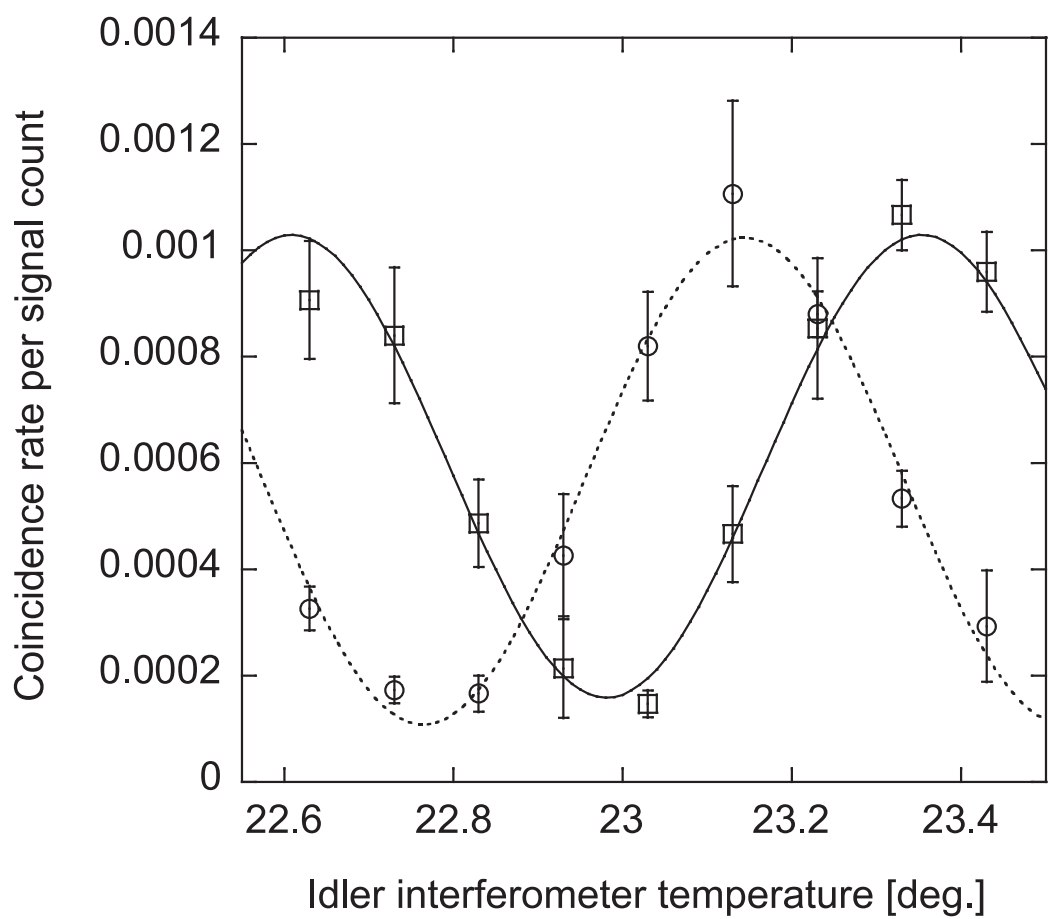}}

\caption{Two-photon interference fringes.}
\label{3}
\end{figure}

We then inserted PLC interferometers. The average photon number per qubit was set at approximately 0.1. First, we fixed the signal interferometer temperature at 12.30 deg., and counted the coincidences while changing the idler interferometer temperature. The result is shown by the circles in Fig. \ref{3} and reveals clear modulation of the coincidence count. The coincidence rate at the peaks of the fringes was about 1.0 Hz. The visibility of the fitted curve was $80.8 \pm 6.3$\%. Note that we have not subtracted the accidental coincidences or detector noise counts in the fringes shown in Fig. \ref{3}. We then changed the signal interferometer temperature to 12.10 deg. to implement a non-orthogonal measurement basis for the signal photons. Here too, we obtained a fringe with $73.2 \pm 7.7$\% visibility. Thus, we confirmed the generation of an entangled state. When the accidental coincidences were subtracted, the visibilities for signal interferometer temperatures of 12.30 and 12.10 deg. were $104.9 \pm 14.1$\% and $94.7 \pm 13.3$\%, respectively. Since the fiber coupling system used in the experiments was desinged for larger waveguides such as a periodically poled lithium niobate waveguide, the coupling between the silicon waveguide and the fiber was not stable for a long time possibly due to the thermal fluctuation of the system. The limited measurement time and the small coincidence rate owing to the relatively large coupling loss resulted in large standard deviations in the visibilities. With an improved fiber coupling system, we will be able to achieve a smaller coupling loss and long-term stability.

Assuming that the number of photon pairs has a Poissonian distribution, which is a valid assumption when the pump pulse width is much larger than the coherence time of the photon pair as in this experiment, the portion of accidental coincidences caused by dark counts (a photon detected in one detector and a dark count in the other, and dark counts in both detectors) is estimated to be approximately 30\%. This means that most of the accidental coincidences were caused by multi-photon-pair emission, which implies that we can further improve visibility by reducing the average photon-pair number.

In conclusion, we have reported entanglement generation using a silicon waveguide. We observed high-purity photon pairs without controlling the temperature of the silicon waveguide, and obtained a time-bin entanglement with $>$73\% visibility. We expect our entanglement source to be useful for multi-node quantum communication systems such as quantum relays. 

The authors thank K. Inoue for fruitful discussions.

\newpage

\newpage

\newpage

\end{document}